\documentclass[twoside]{ilcws07}
\usepackage[latin1]{inputenc}
\usepackage[dvips]{graphicx,epsfig,color}
\usepackage{wrapfig,rotating}
\usepackage{amssymb,amsmath,array}

\begin{document}
\title{
%%%%   Paper title goes here  %%%%%%%%%%%%%%
The ILC Energy Requirements from the Constraints on New Boson
Production at the Tevatron} %%
%***********************************************************************
% AUTHORS INFORMATION AREA
%***********************************************************************
\author{Mihail Chizhov$^{1,2}$
% Optional short acknowledgment: remove next line if non-needed
\thanks{I thank the Local Organisation Committee of the LCWS/ILC07 workshop for the financial support of my participation.}
% DO NOT MODIFY THE FOLLOWING '\vspace' ARGUMENT
\vspace{.3cm}\\
% Addresses and institutions (remove "1- " in case of a single institution)
1- Sofia University, Physics Department \\
BG-1164 Sofia, Bulgaria
%% Remove the next three lines in case of a single institution
\vspace{.1cm}\\
2- H1 Collaboration at DESY \\
Notkestr. 85, D-22607 Hamburg, Germany\\
}
%%***********************************************************************
% END OF AUTHORS INFORMATION AREA
%***********************************************************************

\maketitle

\begin{abstract}
Direct constraints on the masses of new heavy bosons by the Tevatron
data are discussed. Some excesses in the experimental data are
interpreted as a resonance production of new charged and
`leptophobic' neutral chiral bosons with masses around 500~GeV and
700~GeV, respectively. The interpretation was provided on the basis
of the theoretical model,  proposed by the author about 15 years
ago. New Tevatron data and the LHC results will definitely confirm
or reject this interpretation. The ILC with an energy above 1~TeV
would be an ideal place to produce and to study the properties of
these particles.

\end{abstract}

\section{Introduction}

The hadron colliders, due to the biggest center-of-mass energy and
their relatively compact sizes, still remain a main tool for
discoveries of very heavy particles. Thus, in 1983 the two dedicated
experiments UA1~\cite{UA1} and UA2~\cite{UA2} discovered the
intermediate vector bosons at the CERN SPS Collider. One faces,
however, a very large background from the strong interactions.

In any case, besides the simple manifestation of the existence of
the weak bosons, one needs a precise study of their properties
following from the Standard Model (SM). This task has been
excellently fulfilled by the Large Electron-Positron (LEP) storage
ring at CERN and the Stanford Linear Collider (SLC) at SLAC.
However, the masses of the $t$ quark and the undiscovered yet Higgs
boson happened to be too high to be discovered at these colliders.

I remember the words by Samuel Ting at one of the LEP meetings in
defence of continuation of the LEP running: ``Each collision at the
lepton colliders is an event, while it is a background at the hadron
colliders''. So, the precision of the electroweak measurements at
the lepton colliders was so high, that the predicted from the
radiative loop corrections mass of the top-quark
$m_t=180\,^{+8}_{-9}\;^{+17}_{-20}$~GeV~\cite{LEPEWG} has been found
in agreement and with a comparable accuracy of its first direct
measurements at the Fermilab Tevatron by the CDF~\cite{tCDF}
$m_t=176\pm 8\pm 10$~GeV and the D0~\cite{tD0}
$m_t=199\,^{+19}_{-21}\pm 22$~GeV collaborations.

Nevertheless, in spite of the overwhelming background for the
top-quark pair production by the strong interactions at the hadron
collider, the uncertainty of the top-quark mass $m_t=170.9\pm 1.1\pm
1.5$~GeV~\cite{mt} is considerably reduced at present. Moreover,
recently, the evidence for a single top-quark
production~\cite{singleTop} through the weak interactions and the
direct measurement of $|V_{tb}|$ at the Fermilab Tevatron hadron
collider became possible. Another achievement in precise
measurements at the hadron collider is the $W$-mass measurement
$m_W=80.413\pm 0.048$~GeV~\cite{CDFmW} by the CDF collaboration at a
comparable with the LEP experiments accuracy, which represents the
single most precise measurement to date. All these measurements will
allow to constrain further the mass of the Higgs particle, which
discovery is a priority task of the running Tevatron and the Large
Hadron Collider (LHC).

The discovery of the theoretically predicted heavy particles and the
establishment of the SM without any surprises are characteristic for
the experimental high energy physics during the last thirty years.
Therefore, the LHC construction is connected not only with the Higgs
discovery, but with the hope to find the physics beyond the SM. Up
to now it is not clear what kind of physics it will be. Therefore,
any inputs like constraints on the new physics from low-energy
precise experiments or from the presently most powerful Tevatron
collider at FNAL are badly needed when discussing the properties of
future colliders, in particular, the International Linear Collider
(ILC).

This talk is dedicated to the energy requirements for the future
lepton colliders, which follow from the constraints on the new boson
production at the Tevatron. In order to investigate the properties
of the new bosons and eventually to distinguish among different
models of the new physics, the energy of the future ICL should be
enough to produce them. Although it is still possible to investigate
some properties of the new bosons at low-energy, we will consider
the case of their resonance or threshold production, as an optimal
possibility.

In the second part of the talk we will consider one of the possible
scenarios of new physics in the boson sector, for which some
confirmation from the Tevatron data already exists. A quantitative
model of such a new physics will be very valuable in interpreting
the data from the hadron colliders, Tevatron and LHC, that presents
concrete requirements for the ILC energy design.

\section{Tevatron constraints}

Let us start with the case of new neutral massive bosons, $Z'$,
which can be produced at the lepton colliders as resonances. Such a
type of bosons is very difficult to detect in the low-energy
experiments due to the huge background from the electromagnetic
interactions. Some guiding principle is necessary to distinguish
them from the known interactions. For example, the neutral weak
currents were detected in the deep-inelastic electron scattering
through the measurements of $P$-odd quantities. Therefore, we expect
that direct constraints from the high-energy hadron colliders should
be more restrictive.

Moreover, up to now, the Drell-Yan process with high-energy
invariant mass of the lepton pairs remains the most clear indication
of the heavy boson production at the hadron colliders. Therefore,
the constraints from these investigations can be directly applied to
the resonance boson production at the lepton colliders. So, using
only a modest integrated luminosity of 200 pb$^{-1}$ collected
during RUN~II, the D0 Collaboration puts tight restrictions on the
$Z'$ masses for the different models from the di-electron
events~\cite{D0_4375}: $M_{Z'_{SM}} < 780$~GeV, $M_{Z'_\eta} <
680$~GeV, $M_{Z'_\psi} < 650$~GeV, $M_{Z'_\chi} < 640$~GeV and
$M_{Z'_I} < 575$~GeV. A comparable statistics in the di-muon channel
leads approximately to the same constraint $M_{Z'_{SM}} <
680$~GeV~\cite{D0_4577}. The CDF constraints from the di-electron
channel are based on more data, 1.3 fb$^{-1}$, which lead to tighter
restrictions~\cite{CDF_8694}: $M_{Z'_{SM}} < 923$~GeV, $M_{Z'_\eta}
< 891$~GeV, $M_{Z'_\psi} < 822$~GeV, $M_{Z'_\chi} < 822$~GeV and
$M_{Z'_I} < 729$~GeV.

\begin{figure}[ht]
\hspace{1cm}\includegraphics[width=0.5\columnwidth]{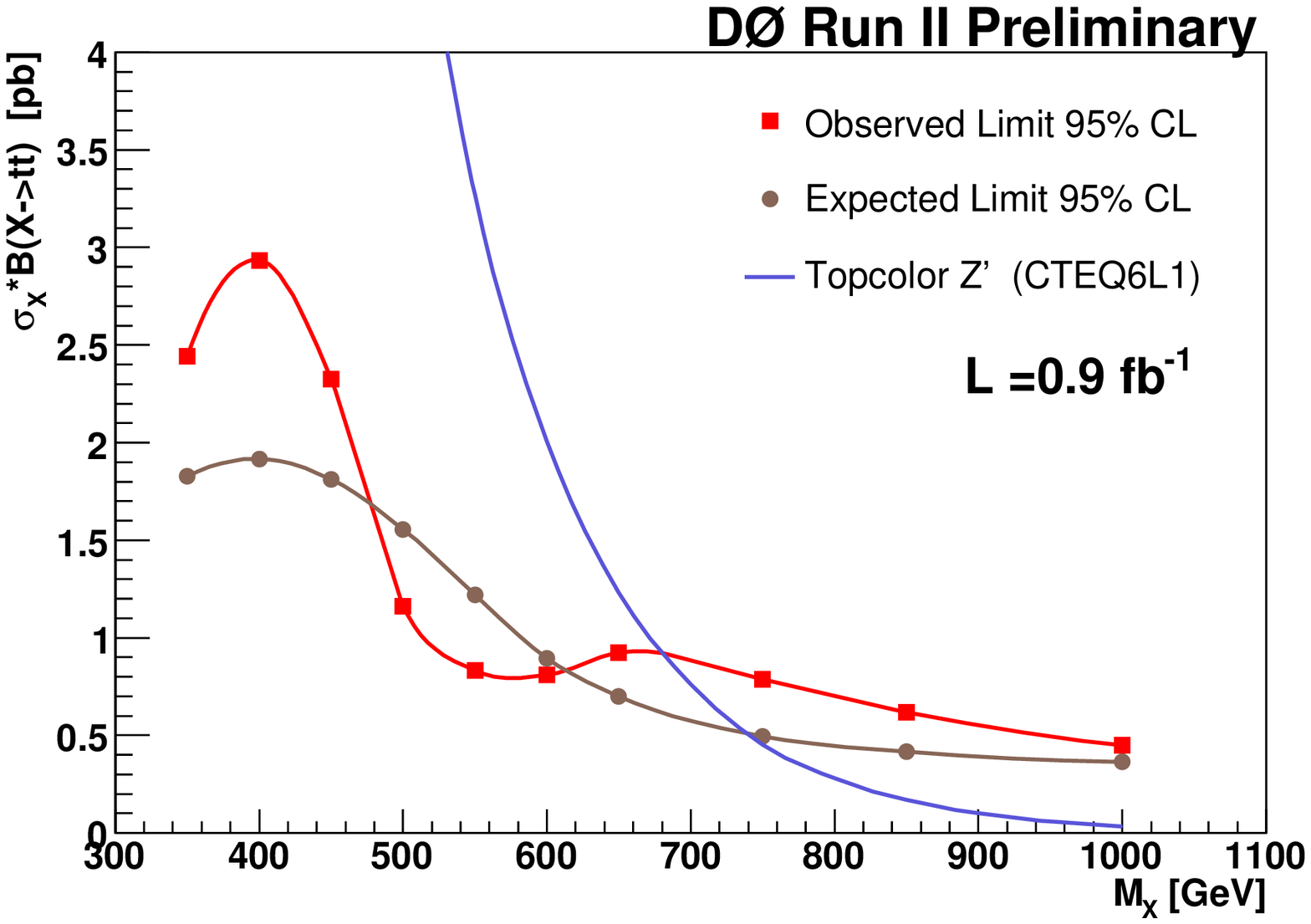}
\includegraphics[width=0.36\columnwidth]{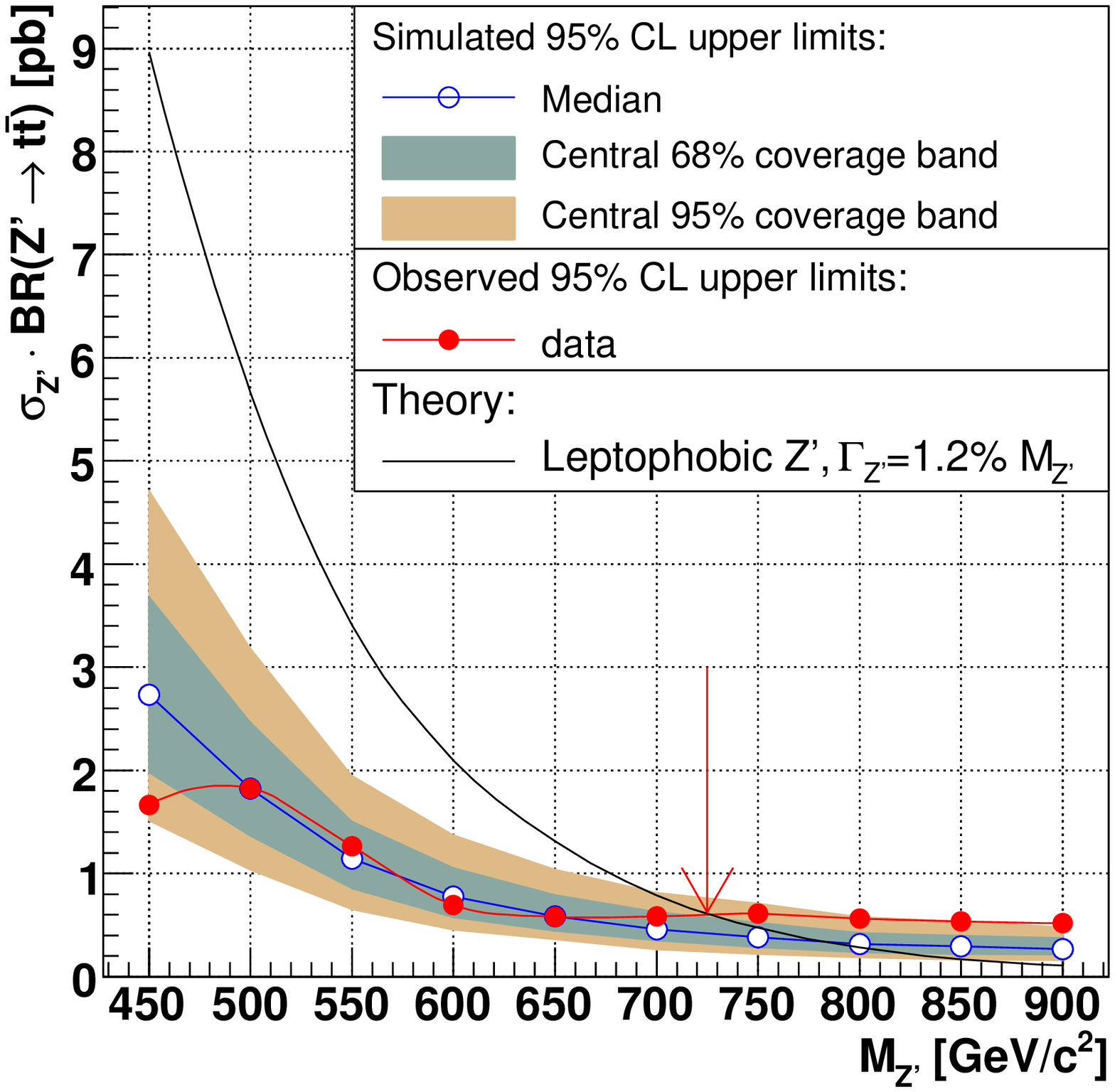}
\caption{Expected and observed 95\% C.L. upper limits on
$\sigma_X\times{\cal B}(X\to t\bar{t})$ in comparison with the
predicted leptophobic topcolor $Z'$ cross section (left panel from
\cite{D0ttbar} -- D0 data, right panel from \cite{CDFttbar} -- CDF
data of 680 pb$^{-1}$).}\label{Fig:Mttbar}
\end{figure}

Another possible channel, which can indicate the production of the
neutral heavy bosons, is their hadronic decay into $t\bar{t}$ pairs.
While the light quark decay channels are swamped by multijet
background, the $t\bar{t}$ pairs can be detected, for example,
through their decays into two energetic $b$-jets and two $W$'s,
where one $W$ boson decays hadronically and one leptonically.
Although the constraints from this channel cannot be applied
directly to the energy requirements for the lepton collider due to
the possible leptophobic character of the bosons, it is interesting
to detect the eventual peaks in the Tevatron data. So the latest
results both of the D0~\cite{D0ttbar} and of the CDF~\cite{CDFttbar}
Collaborations show some excess in the invariant mass distributions
around 700~GeV (Fig.~\ref{Fig:Mttbar}). A possible explanation of
this excess will be discussed in the next section.

Let us consider the case of the new heavy charged bosons,
generically noted by $W'$. They could be produced at the lepton
colliders only in pairs or in association with other charged boson,
like $W$. Therefore, restrictions on their masses lead to the
following energy requirements for their threshold production $E >
M_{W'}+M_W$ at the lepton colliders. Here again we will consider
leptonic and hadronic channels of their decays.

The leptonic decay of the new heavy charged boson into high-energy
pair of a lepton and a corresponding antineutrino is the most clear
signature of its production at the hadron colliders. So, already
from 205 pb$^{-1}$ of RUN~II data, the CDF Collaboration obtained a
tight constraint on possible $W'$ mass $M_{W'} >
788$~GeV~\cite{WpCDFlepton}. The most rigid constraint comes from
the D0 Collaboration~\cite{WpD0lepton} $M_{W'} > 965$~GeV, based on
bigger statistics, 900 pb$^{-1}$, and better calorimetry than the
CDF detector.

The hadronic decay of the new heavy charged boson into a $t\bar{b}$
pair of a heavy $b$ quark and a short living $t$ quark with its
subsequent decay to $Wb$ pair allows to make jet $b$-tagging, where
one of the jets must have a displaced secondary vertex. A search for
the intermediate heavy bosons in this channel has been fulfilled by
both the D0 and CDF collaborations, and for this purpose the part of
the same data sets of the single top production analysis has been
used. Owing to boson high masses this analysis is even simpler than
the single top production searches, because at such energies the
background is considerably reduced.

\begin{wrapfigure}{r}{0.5\columnwidth}
\centerline{\includegraphics[width=0.42\columnwidth]{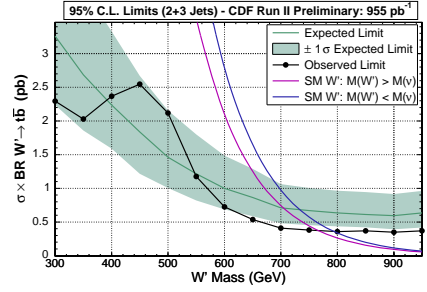}}
\caption{Observed limits from \cite{WpCDFquark}.}\label{Fig:23j_obs}
\end{wrapfigure}

So the D0 Collaboration, based on 230 pb$^{-1}$ of integrated
luminosity, puts the following constraints on the $W'$ mass
depending on the model: $M_{W'_{SM}} > 610$~GeV,
$M_{W'_R(\to\,\ell~{\rm and}~q)} > 630$~GeV, and $M_{W'_R(\to\,
q~{\rm only})} > 670$~GeV~\cite{WpD0quark}. The CDF constraints are
tighter (Fig.~\ref{Fig:23j_obs}): $M_{W'} > 760$~GeV for $M_{W'} >
M_{\nu_R}$ and $M_{W'} > 790$~GeV for $M_{W'} < M_{\nu_R}$, since
they are
based on 955 pb$^{-1}$~\cite{WpCDFquark}.

Taking into account the most stringent constraints from the Tevatron
data, we can conclude, that in order to produce the heavy charged
boson in association with the $W$ boson or the heavy neutral boson a
lepton collider with energy above 1~TeV is necessary. Also it is
interesting to note the presence of some excesses in the observed
data in Figs.~\ref{Fig:Mttbar} and \ref{Fig:23j_obs}, which we will
discuss in the next section.

\section{New spin-1 chiral bosons}

Additional chiral bosons, which have anomalous interactions with
fermions, were proposed in \cite{MPL}. An exchange through these
bosons leads to effective tensor interactions with the coupling
constant by two orders of magnitude smaller than $G_F$. This follows
from the precise low-energy experiments of the radiative pion
decay~\cite{RPD}. Assuming the universality of these interactions we
can explain the long standing discrepancy between the two pion
production in the $e^+e^-$ annihilation and the $\tau$
decay~\cite{tau}, which now reaches 4.5$\,\sigma$~\cite{Davier}.

The universality of the interactions of the new bosons and the
hypothesis about a dynamical generation of their kinetic terms allow
to predict their masses~\cite{production}. Due to the mixing between
two charged chiral bosons the lightest state corresponds to
$U^\pm$-boson with a mass $M_L\approx 509$~GeV and the heaviest one
is $T^\pm$-boson with a mass $M_H\approx 1137$~GeV. The neutral
physical states come as $CP$-even $U^R$ and $CP$-odd $U^I$ bosons
with approximately
\begin{wrapfigure}{r}{0.5\columnwidth}
\centerline{\includegraphics[width=0.42\columnwidth]{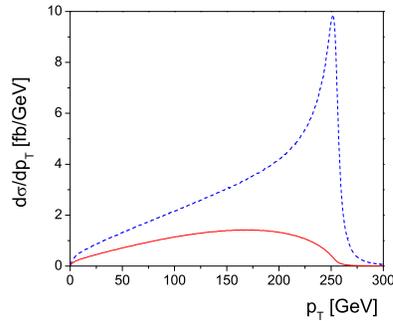}}
\caption{The distributions for the gauge $W'$ (dashed) and for the
chiral $U^\pm$ (solid) bosons as functions of the lepton transfers
momentum.}\label{Fig:3}
\end{wrapfigure}
the same masses $M_U\approx 719$~GeV, which couple only to the up
fermions, and analogous but heavier bosons $T^R$ and $T^I$ with a
common mass $M_T\approx 1017$~GeV, coupling to the down fermions.

\begin{wrapfigure}{r}{0.5\columnwidth}
\centerline{\includegraphics[width=0.42\columnwidth]{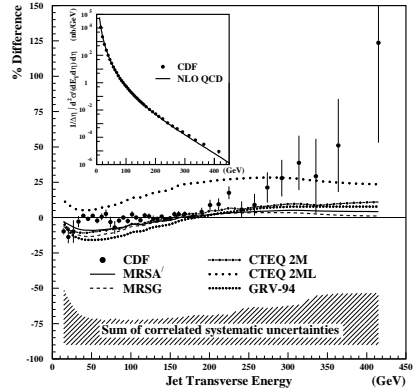}}
\caption{The Fig.~1 from \cite{CDFjet}}\label{Fig:4}
\end{wrapfigure}

Due to the anomalous interactions the angular distribution of the
chiral boson decays differs drastically from the analogous
distribution of the gauge bosons. This leads to a specific
transverse momentum distribution~\cite{TevLHC}, which has a broad
smooth bump with a maximum below the kinematical endpoint $p_T=M/2$,
instead of a sharp Jacobian peak (Fig.~\ref{Fig:3}). The form of the
decay distribution for the chiral bosons resembles the bump
anomalies in the inclusive jet $E_T$ distribution
(Fig.~\ref{Fig:4}), reported by the CDF Collaboration~\cite{CDFjet}
many years ago.

Analysing the bumps in the jet transverse energy distribution in
Fig.~\ref{Fig:4}, we can find the endpoint of the first bump at
250~GeV and guess about the second bump endpoint from the minimum
around 350~GeV. If we assign the first bump to the hadron decay
products of the lightest charged bosons, which exactly corresponds
to the estimated mass, the second endpoint hints to a mass around
700~GeV, which is also in a quantitative agreement with our
estimations for the mass of the lightest neutral boson. However,
taking into account the large systematic uncertainties in jet
production, these conclusions may be premature, unless they are
confirmed in other channels.

Indeed, an excess about 2$\sigma$ in the lepton channel has been
pointed out recently by the CDF Collaboraion~\cite{WpCDFlepton} in
the region $350$~GeV $<M_T\simeq 2p_T<500$ GeV. At the same time the
same collaboration, however, denies the peak in the quark channel in
the same region (Fig.~\ref{Fig:23j_obs}), claiming that ``since the
predictions in {\em the neighboring bins} agree with the
observation, and since {\em the three jet bin} does not show a
similar excess, we anticipate that the excess in this region is a
statistical fluctuation''. But this signature means just a resonance
and this excess is in some sense a confirmation of the excess in the
leptonic channel!

Therefore, the independent result from the D0 collaboration is very
important. Their published result~\cite{WpD0quark} is based on 230
pb$^{-1}$ of integrated luminosity and does not show any excess in
the histogram with the bin's width of 50 GeV. However, it should
always be taken into account that the narrow peak could be missed
due to the smearing effect of the detector resolution or an
insufficient statistics. Indeed, the right histogram in the Fig.~3
of the conference paper \cite{WpD0quarkNote} of the same
collaboration with the bin's width of the 45 GeV reveals,
nevertheless, the weak peak in the same region of the 500 GeV. All
these not statistically significant results for the separated
analyses may give a more conclusive answer after their combination
and an additional investigation of the angular distributions of the
events in this region.

The small excess in the $t\bar{t}$ channel around 700~GeV
(Fig.~\ref{Fig:Mttbar}) can be explained in the framework of our
model by the production and the decay of the lightest neutral chiral
boson. The latter shows `leptophobic' property, since it decays to
`invisible' $\nu\bar{\nu}$ leptonic channel, and can be detected
only through its decay into a pair of up quarks. The D0
Collaboration even superimposed its plot of the $t\bar{t}$ invariant
mass distribution with the expected signal for a topcolor-assisted
technicolor $Z'$ with $M_{Z'} = 750$~GeV, which perfectly agrees
with the data.

\section{Conclusions}

\begin{figure}[ht]
\hspace{0.5cm}\includegraphics[width=0.45\columnwidth]{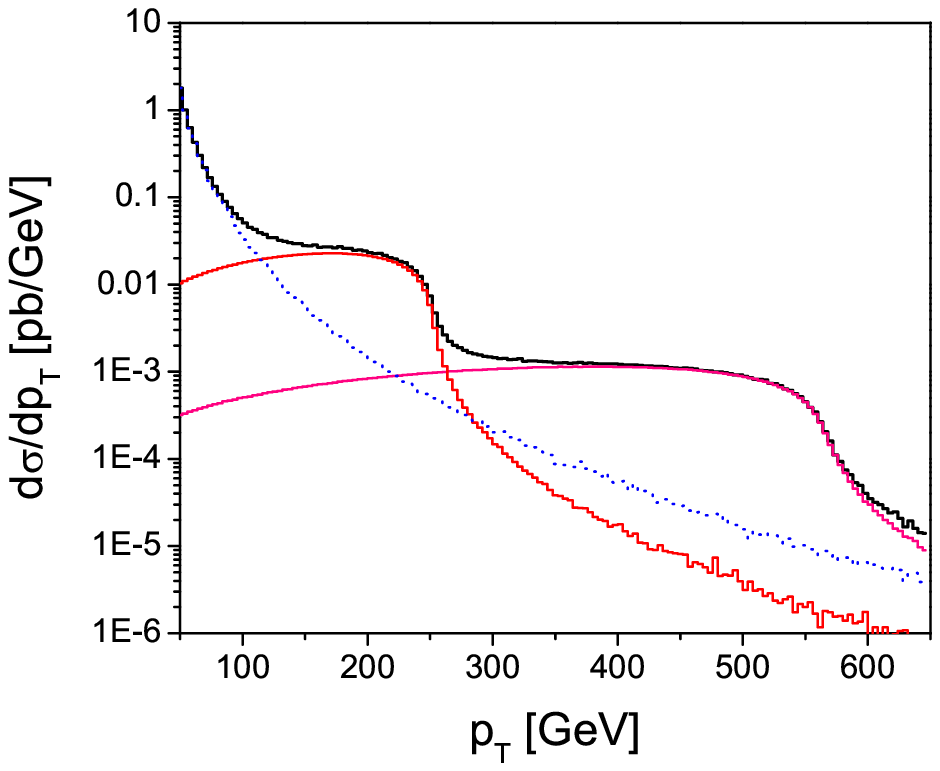}
\includegraphics[width=0.45\columnwidth]{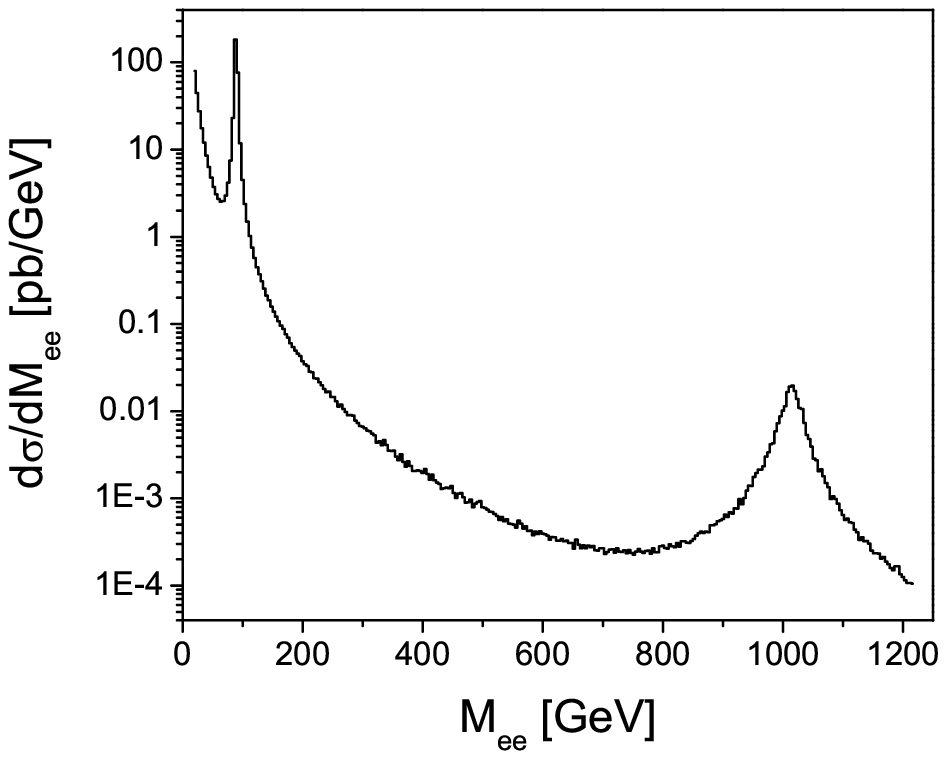}
\caption{The distributions in the lepton channels at the LHC, namely
$pp\to eE\!\!\!\!/_T$ (left) and $pp\to e^+ e^-$
(right).}\label{Fig:LHC}
\end{figure}

There are some hints for the existence of a lightest charged chiral
boson with a mass around 500~GeV and a lightest neutral
`leptophobic' chiral boson with a mass around 700~GeV in the
Tevatron data. In the positive case the LHC would be able to
discover all predicted charged and neutral chiral bosons spanning in
mass up to around 1 TeV (see their leptonic decay distributions in
the Fig.~\ref{Fig:LHC}). The ILC with such energy would be an ideal
place to produce and to study these particles.

% ****************************************************************************
% BIBLIOGRAPHY AREA
% ****************************************************************************

\begin{footnotesize}
% IF YOU DO NOT USE BIBTEX, USE THE FOLLOWING SAMPLE SCHEME FOR THE REFERENCES
% ----------------------------------------------------------------------------

\end{footnotesize}

% ****************************************************************************
% END OF BIBLIOGRAPHY AREA
% ****************************************************************************

\end{document}